\documentclass[10pt, final, journal, letterpaper, twocolumn]{IEEEtran}
\makeatletter
\def\ps@headings{%
	\def\@oddhead{\mbox{}\scriptsize\rightmark \hfil \thepage}%
	\def\@evenhead{\scriptsize\thepage \hfil \leftmark\mbox{}}%
	\def\@oddfoot{}%
	\def\@evenfoot{}}
\makeatother 

\IEEEoverridecommandlockouts
\usepackage{bbm}
\usepackage{amsfonts}
\usepackage[dvips]{graphicx}
\usepackage{times}
\usepackage{cite}
\usepackage{amsmath}
\usepackage{array}
\usepackage{amssymb}

\usepackage{stfloats}
\usepackage{graphicx}
\usepackage{footnote}
\usepackage{booktabs}
\usepackage{array}
\usepackage{algorithm}
\usepackage{subeqnarray}
\usepackage{cases}
\usepackage{threeparttable}
\usepackage{color}
\usepackage{hyperref}
\usepackage{epstopdf}
\usepackage{algpseudocode}
\usepackage{bm}
\usepackage{multirow}
\usepackage[labelformat=simple]{subcaption}
\usepackage{adjustbox}

\allowdisplaybreaks[4]

\usepackage{geometry}
\begin{document}
	
	\title{Atomic Norm Minimization-based DoA Estimation for IRS-assisted Sensing Systems}

	\author{\authorblockN{Renwang Li, Shu Sun, \IEEEmembership{Member,~IEEE}, and Meixia Tao, \IEEEmembership{Fellow,~IEEE}}\\
		\thanks{This work is supported by the Natural Science Foundation of China under Grant 62125108 and Grant 62271310, and by the Fundamental Research Funds for the Central Universities of China. The authors are with the Department of Electronic Engineering and the Cooperative Medianet Innovation Center (CMIC), Shanghai Jiao Tong University, China (email:\{renwanglee, shusun, mxtao\}@sjtu.edu.cn). (Corresponding authors: Shu Sun and Meixia Tao)}
	}
	
	\maketitle
	
	\vspace{-1.5cm}
	
	\begin{abstract}
		Intelligent reflecting surface (IRS) is expected to play a pivotal role in future wireless sensing networks owing to its potential for high-resolution and high-accuracy sensing. In this work, we investigate a multi-target direction-of-arrival (DoA) estimation problem in a semi-passive IRS-assisted sensing system, where IRS reflecting elements (REs) reflect signals from the base station to targets, and IRS sensing elements (SEs) estimate DoA based on  echo signals reflected by the targets. {First of all, instead of solely relying on IRS SEs for DoA estimation as done in the existing literature, {this work fully exploits  the DoA information embedded in both IRS REs and SEs matrices via the atomic norm minimization (ANM) scheme.} Subsequently, the Cram\'er-Rao bound  for DoA estimation is derived, revealing an inverse proportionality to $MN^3+NM^3$ under the case of identity covariance matrix of the IRS measurement matrix and a single target, where $M$ and $N$ are the number of IRS SEs and REs, respectively. Finally, extensive numerical results substantiate the superior accuracy and resolution performance of the proposed ANM-based DoA estimation method over  representative baselines.}
	\end{abstract}
	
	\section{Introduction}
	The future sixth-generation (6G) wireless networks are anticipated to require high-resolution and high-accuracy sensing to support environment-aware applications such as autonomous driving \cite{9349624}. {However, current cellular base station (BS)-based sensing encounters challenges, including   dead zones for sparsely deployed BSs,   blockage issue for obstructed obstacles, and   severe path loss for round-trip propagation in millimeter-wave systems.}
	
	To address these challenges, intelligent reflecting surface (IRS) is regarded as a promising technology\cite{9122596}. Generally, IRS is composed of many passive reflecting elements (REs) which can independently impose phase shifts on the incident signals. {It can create a non-line-of-sight (NLoS) link for   dead zones or blocked areas, thus ensuring uninterrupted sensing.} Additionally, IRS can provide high beamforming gain due to its large aperture, thus extending the sensing coverage. Consequently, IRS-aided sensing systems are widely studied \cite{9508872, 9695358, 9954622}. For instance, the authors in \cite{9954622} design the measurement matrix of IRS and adopt an efficient method for target angle estimation.
	
	However, the severe path loss of the IRS-aided sensing systems  limits their application dramatically. The path loss of the BS-IRS-target-IRS-BS link is much higher than that of the conventional BS-target-BS link. Therefore, semi-passive IRS, composed of passive REs and active sensing elements (SEs), is proposed to address the path loss issue  \cite{9724202}.   IRS SEs are employed to collect   echo signals reflected from   targets and directly sense them, thus reducing   path loss by avoiding the IRS-BS link. Some researchers have explored   semi-passive IRS-aided sensing systems \cite{song2023fullypassive, 10304548,10284917}. The authors in \cite{song2023fullypassive} compare the detection signal-to-noise ratio and estimation Cram\'er-Rao bound (CRB) performance of fully-passive and semi-passive IRS-enabled sensing systems, and reveal the outstanding performance of semi-passive IRS except for   scenarios with an  extremely large number of REs. The authors in \cite{10304548} propose a new beam scanning-based protocol for simultaneously target sensing and user communication. The authors in \cite{10284917} consider a joint time-of-arrival and direction-of-arrival (DoA) estimation problem, and adopt the multiple signal classification (MUSIC) method for DoA estimation. However,   IRS not only establishes a NLoS link for obstructed targets, but also measures   targets with IRS phase offsets. The MUSIC method relies solely on the receive antenna structure, and fails to fully exploit the antenna resources in the semi-passive IRS-aided sensing system.   More specifically, in addition to SEs, the target angle information is also embedded in IRS REs, a factor not considered in \cite{10284917}. Therefore, the DoA estimation performance can be significantly enhanced by further leveraging the DoA information from IRS REs.
	
	\begin{figure}[t]
		\begin{centering}
			\includegraphics[width=.38\textwidth]{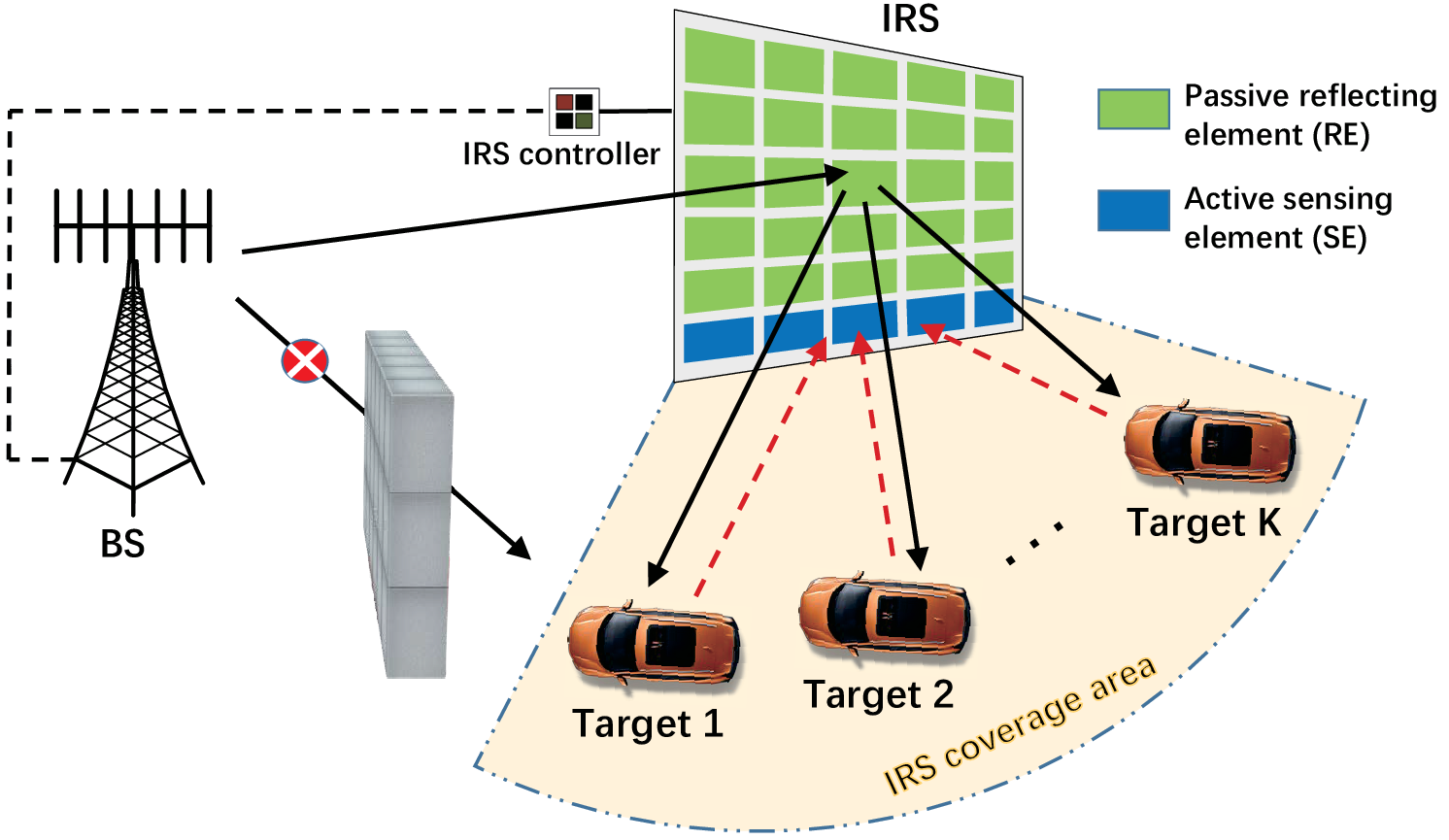}
			\caption{{Semi-passive IRS assisted localization system.}}\label{fig_system}
		\end{centering}
		\vspace{-0.9cm}
	\end{figure}	
	
	Based upon the  background above, we address a multi-target DoA estimation problem in a semi-passive IRS-assisted sensing system as illustrated in Fig. \ref{fig_system}. {First, {in order to fully exploit the DoA information in both IRS REs and SEs matrices}, a reconstruction problem is proposed by utilizing the atomic norm minimization (ANM) method to reconstruct a sparse signal  composed of finite atoms from the received echo signal.} The original ANM problem is intractable and is thus transformed into a convex form according to the Lagrangian dual theorem. Next, the CRB for DoA estimation is   derived to characterize the estimation performance. When the covariance matrix of the IRS measurement matrix is an identity matrix, the CRB is inversely proportional to $MN^3+NM^3$ for the single-target case, where $M$ and $N$ are the numbers of IRS SEs and REs, respectively. Finally, extensive numerical results validate the superior accuracy and resolution performance of the proposed ANM-based DoA estimation method over the MUSIC method. Furthermore, when the measurement matrix of IRS REs is configured as a discrete Fourier transform (DFT) matrix, the mean square error (MSE) of the proposed method closely approximates the CRB.
	
	\section{System Model}
	We consider a localization system with the assistance of a semi-passive IRS as illustrated in Fig. \ref{fig_system}, where an $N_b$-antenna BS attempts to detect the angles of $K$ targets. The direct links between the BS and targets are assumed to be blocked due to an unfavorable propagation environment. Thus, an IRS is deployed to establish NLoS links for target angle estimation. Here, a semi-passive IRS consisting of  $N$ passive REs  and $M$ active SEs  is adopted, which offers better sensing performance compared to the fully-passive IRS \cite{9724202}. We consider a quasi-static flat-fading channel within the considered time slots.
	
	\vspace{-0.5 cm}
	\subsection{Channel Model}
	For ease of exposition, the uniform linear array (ULA) structure is considered at the BS, IRS REs, and IRS SEs, which can be extended to other array structures. We consider the line-of-sight channel model to characterize the BS-IRS REs channel $\mathbf{H}_{BI}$,
	\begin{equation} 
		\mathbf{H}_{BI} = \alpha_g \mathbf{a}_r ( \theta_{BI}) \mathbf{a}_b^H (\vartheta_{BI}),
	\end{equation}	
	where {$\alpha_g=\frac{\lambda}{4\pi d_{BI}} e^{\frac{-j 2\pi d_{BI}}{\lambda}}$} denotes the complex-valued path gain, $\lambda$ is the carrier wavelength, $d_{BI}$ is the distance between the BS and IRS,  $\vartheta_{BI}$ denotes the angle of departure from the BS, $\theta_{BI}$ denotes the angle of arrival  to the IRS, and $\mathbf{a}_r(\cdot)$ $\left(\mathbf{a}_b(\cdot)\right)$ represents the array response vector of the IRS REs (BS). The array response vector for a ULA with $N$ elements of half-wavelength spacing can be expressed as
	\begin{equation} \label{exp_array_resp}
	\hspace{-0.25cm}	{ {\mathbf{a}}(\theta) = \left[ 
		1 \enspace e^{-j \pi \sin (\theta)}  \enspace \dots \enspace e^{-j (N-1)\pi \sin (\theta)}
		\right]^{T} \in \mathbb{C}^{N\times 1}.}
	\end{equation}
    The IRS REs-target $k$-IRS SEs channel can be expressed as
	\begin{equation} 
		\mathbf{H}_k = \alpha_k \mathbf{b} ( \theta_{k}) \mathbf{a}_r^H (\theta_{k} ),
	\end{equation}
	where  $\theta_{k}$ denotes the DoA of target $k$, $\mathbf{b}(\cdot)$  represents the array response vector of the IRS  SEs,  {$\alpha_k=\sqrt{\frac{\lambda^2 \kappa_k}{64\pi^3 d_{k}^4}} e^{\frac{-j 4\pi d_{k}}{\lambda}}$} \cite{9367457} refers to the path gain of the IRS REs-target $k$-IRS SEs link, $d_{k}$ denotes the  distance between the IRS and target $k$, and $\kappa_k$ is the radar cross section (RCS) of target $k$.
	
	\vspace{-0.5 cm}
	\subsection{IRS-assisted Sensing Model}
	The IRS SEs concurrently  receive  both the direct signals from the BS and echo signals reflected by the targets. However, the BS-IRS SEs signals do not carry  target information, and can be perfectly pre-canceled at the IRS SEs. This is feasible because the fixed and known locations of the BS and IRS allow for the pre-estimation of the BS-IRS channel \cite{10304548}.  Consequently, only the echo signals are considered at the IRS SEs. The received echo signals $\mathbf{y}(t) \in \mathbb{C}^{M\times 1}$ at the IRS SEs in time slot $t$ are given by
	\vspace{-0.2cm}
	\begin{equation} \label{y_rece}
		{\mathbf{y}(t) = \sqrt{P_t} \sum_{k=1}^K \mathbf{H}_k \operatorname{diag} (\boldsymbol{\phi}) \mathbf{H}_{BI} \mathbf{w} s(t)   + \mathbf{n}(t),}
	\end{equation}	
	where $P_t$ is the transmit power at the BS, $s(t)$ is the transmit signal with unit power, $\mathbf{w} \in \mathbb{C}^{N_b\times 1}$ is the transmit beamforming vector with $\|\mathbf{w}\|^2=1$, $\mathbf{n}(t) \sim \mathcal{CN}(0, \sigma^2 \mathbf{I}_M)$ is the receiver additive white Gaussian noise, and $\boldsymbol{\phi}=\left[e^{j\phi_1}, e^{j\phi_2}, \ldots, e^{j\phi_N}\right]^T$ denotes the reflection vector of IRS REs \footnote{{For simplicity, the RCS of IRS elements is assumed to be 1 \cite{9508872, 9695358, 9954622, 9724202, song2023fullypassive, 10304548,10284917}. More detailed RCS models can be found in \cite{9732186}}}. 
	
	As the BS-IRS REs channel $\mathbf{H}_{BI}$ can be pre-estimated at the IRS SEs, the optimal transmit beamforming is obtained as $\mathbf{w}=\frac{1}{\sqrt{N_b}} \mathbf{a}_b (\vartheta_{BI})$. {The IRS reflection vector remains constant within each time slot but varies between different time slots to obtain diverse measurements of the targets for DoA estimation.} The received signals can be rewritten as \cite{10304548}
	\begin{align}  
		\mathbf{y}(t) = \sqrt{N_b P_t} \alpha_g \sum_{k=1}^K \alpha_k \mathbf{b}(\theta_k) \mathbf{a}_r^H (\overline{\theta}_k) \boldsymbol{\phi}(t) s(t)  + \mathbf{n}(t),
	\end{align}	
	where $\overline{\theta}_k = \arcsin (\sin(\theta_k) - \sin(\theta_{BI}))$. Collecting these received signals during $L$ time slots, yielding
	\begin{align} \label{Y_re}
		\mathbf{Y} \triangleq & \left[\mathbf{y}(1), \mathbf{y}(2), \ldots, \mathbf{y}(L)\right] \in \mathbb{C}^{M\times L} \notag \\
		= &\sqrt{N_b P_t} \alpha_g \sum_{k=1}^K \alpha_k \mathbf{b}(\theta_k) \mathbf{a}_r^H (\overline{\theta}_k) \notag \\
		&\times [\boldsymbol{\phi}(1) s(1), \boldsymbol{\phi}(2) s(2), \ldots, \boldsymbol{\phi}(L) s(L)]   + \mathbf{N} \notag\\
		\triangleq &\mathbf{B}(\boldsymbol{\theta}) \boldsymbol{\Lambda} \mathbf{Q}^H (\boldsymbol{\theta}) \mathbf{D} + \mathbf{N},
	\end{align} 
	where $\mathbf{N} \triangleq [\mathbf{n}(1), \dots, \mathbf{n}(L ) ]\in \mathbb{C}^{M\times L} $, $\boldsymbol{\theta} \triangleq [\theta_1, \ldots, \theta_K]^T \in \mathbb{R}^{K\times 1}  $, $\mathbf{B}(\boldsymbol{\theta}) \triangleq [ \mathbf{b}(\theta_1), \mathbf{b}(\theta_2), \ldots, \mathbf{b}(\theta_K)] \in \mathbb{C}^{M\times K}  $, $\boldsymbol{\Lambda} \triangleq \sqrt{N_b P_t} \alpha_g \operatorname{diag} (\alpha_1, \alpha_2, \ldots, \alpha_K) \in \mathbb{C}^{K\times K}  $, $\mathbf{Q}(\boldsymbol{\theta}) \triangleq [ \mathbf{a}_r(\overline{\theta}_1), \mathbf{a}_r(\overline{\theta}_2), \ldots, \mathbf{a}_r(\overline{\theta}_K)] \in \mathbb{C}^{N\times K}  $,  and $\mathbf{D} \triangleq [\boldsymbol{\phi}(1) s(1), \ldots, \boldsymbol{\phi}(L) s(L)]  \in \mathbb{C}^{N\times L}  $ denotes the measurement matrix of the IRS REs. {Note that the measurement matrix can be predefined at the BS  and  transferred to the IRS via the IRS controller in advance to ensure accurate DoA estimation at the IRS SEs.}
	
	The IRS-assisted sensing model is expressed as \eqref{Y_re}. Our goal is to accurately estimate the DoAs  of the targets based on the received signal $\mathbf{Y}$ and the measurement matrix $\mathbf{D}$, with unknown $\mathbf{B}(\boldsymbol{\theta})$, $\boldsymbol{\Lambda}$, and $\mathbf{Q}(\boldsymbol{\theta})$. 
	
	\section{ANM-Based DoA Estimation Method}
	In order to estimate the target angles, the MUSIC method is employed in \cite{10284917}. However, the MUSIC method is solely dependent on the receive antenna structure, which means that it only utilizes the information of $\mathbf{B}(\boldsymbol{\theta})$ for DoA estimation.   In the semi-passive IRS-assisted sensing model, DoA information is embedded in both the IRS SEs matrix $\mathbf{B}(\boldsymbol{\theta})$ and the REs matrix $\mathbf{Q}(\boldsymbol{\theta})$, marking a significant departure from the conventional sensing model where DoA information is typically confined to the receive antenna matrix $\mathbf{B}(\boldsymbol{\theta})$. Therefore, an ANM-based DoA estimation method is proposed in this work to fully exploit 
	$\mathbf{B}(\boldsymbol{\theta})$ and $\mathbf{Q}(\boldsymbol{\theta})$.
	
	ANM is a well-known infinite-size dictionary based compressive sensing approach, which is widely used in the DoA estimation problem \cite{9695358, 9954622}.  For a matrix $\mathbf{X} \in \mathbb{C}^{M\times N}$, the atomic norm of $\mathbf{X}$ is defined as
	\vspace{-0.2cm}
	\begin{align} \label{exp_atomic}
		\|\boldsymbol{X}\|_{\mathcal{A}} \triangleq &\inf \bigg\{\sum_i c_i: \boldsymbol{X}=\sum_i c_i e^{j \varphi_i} \mathbf{b}(\theta_i) \mathbf{a}_r^H(\overline{\theta}_i) , \notag \\
		&c_i>0, \varphi_i \in[0,2 \pi), \theta_i \in\left(-\frac{\pi}{2}, \frac{\pi}{2}\right] \bigg\},
	\end{align}
	where $c_i\in\mathbb{R}$ describes the atomic decomposition and $\varphi_i$ denotes the corresponding phase. {The received signal $\mathbf{Y}$ is composed of $K$ atoms of $\mathbf{b}(\theta_k) \mathbf{a}_r^H(\overline{\theta}_k)$, and is thus sparse in the domain constructed by the atoms $\mathbf{b}(\theta_i) \mathbf{a}_r^H(\overline{\theta}_i)$. Consequently, we aim to reconstruct a sparse signal $\mathbf{X}$ from the received echo signal $\mathbf{Y}$, yielding the following reconstruction problem }
	\begin{equation} \label{prob_ori}
		\min \limits_{ \mathbf{X} } \quad  \frac{1}{2} \left\| \mathbf{Y} - \mathbf{XD} \right\|_F^2 + \beta \|\mathbf{X}\|_{\mathcal{A}},
	\end{equation}
	where $\beta$ is a parameter balancing the sparsity and   reconstruction performance,  and $\|\mathbf{X}\|_{\mathcal{A}}$ quantifies the sparsity of $\mathbf{X}$. The  problem \eqref{prob_ori} is intractable and challenging to solve. Therefore, we  transform it into a tractable problem below.
	
	First, the original problem \eqref{prob_ori} can be rewritten as
	\begin{subequations} \label{prob_z}
		\begin{align}
			\min \limits_{ \mathbf{X}, \mathbf{Z} } \quad & \frac{1}{2} \left\| \mathbf{Y} - \mathbf{XD} \right\|_F^2 + \beta \|\mathbf{Z}\|_{\mathcal{A}} \\
			\text { s.t. } \quad & \mathbf{Z} = \mathbf{X}.
		\end{align}
	\end{subequations}
	By introducing a Lagrangian parameter $\mathbf{G}\in \mathbb{C}^{M\times N}$, the corresponding Lagrangian function is given by
	\begin{equation}
		\mathcal{L}(\mathbf{X},\mathbf{Z},\mathbf{G}) \triangleq \frac{1}{2} \left\| \mathbf{Y} - \mathbf{XD} \right\|_F^2 + \beta \|\mathbf{Z}\|_{\mathcal{A}} + \langle \mathbf{X}-\mathbf{Z}, \mathbf{G} \rangle.
	\end{equation}
	where $\langle \mathbf{X}-\mathbf{Z}, \mathbf{G} \rangle \triangleq \Re\left\{ \operatorname{Tr}\left(\mathbf{G}^H (\mathbf{X} - \mathbf{Z}) \right) \right\}$. Thereby, the dual function can be obtained as
	\begin{align}
		g(\mathbf{G}) \triangleq & \inf_{\mathbf{X},\mathbf{Z}} \; \mathcal{L}(\mathbf{X},\mathbf{Z},\mathbf{G}) \notag \\
		= & \inf_{\mathbf{X}} \left(\frac{1}{2} \left\| \mathbf{Y} - \mathbf{XD} \right\|_F^2+ \langle\mathbf{X},\mathbf{G} \rangle \right) \notag\\
		&+ \inf_{\mathbf{Z}} \left(\beta \|\mathbf{X}\|_{\mathcal{A}} -\langle\mathbf{Z},\mathbf{G} \rangle \right) \notag\\
		= & \frac{1}{2} \Big\{  - \operatorname{Tr}\left[(\mathbf{Y}\mathbf{D}^H- \mathbf{G})(\mathbf{D} \mathbf{D}^H)^{-1} \left( \mathbf{Y}\mathbf{D}^H- \mathbf{G}\right)^H \right] \notag\\
		& + \operatorname{Tr}(\mathbf{Y}\mathbf{Y}^H) \Big\} -  I_{\mathbf{G}:\|\mathbf{G}\|_{\tilde{\mathcal{A}}} \leq \beta} (\mathbf{G}),
	\end{align}
	where $\|\mathbf{G}\|_{\tilde{\mathcal{A}}}$ is the dual norm of the atomic norm, defined as $\|\mathbf{G}\|_{\tilde{\mathcal{A}}} \triangleq \sup_{\|\mathbf{U}\|_{\mathcal{A}} \leq 1} \langle\mathbf{U}, \mathbf{G}\rangle $, and $I_S(x)$ is an indicator function that equals zero if $x\in S$ and infinity otherwise.
	Hence, the optimization problem \eqref{prob_z} can be transformed into
	\begin{subequations} \label{prob_dual}
		\begin{align}
			\min \limits_{ \mathbf{G}} \quad & \operatorname{Tr}\left[(\mathbf{Y}\mathbf{D}^H- \mathbf{G})(\mathbf{D} \mathbf{D}^H)^{-1} \left( \mathbf{Y}\mathbf{D}^H- \mathbf{G}\right)^H \right] \\
			\text { s.t. } \quad & \|\mathbf{G}\|_{\tilde{\mathcal{A}}} \leq \beta.
		\end{align}
	\end{subequations}
	The dual norm $\|\mathbf{G}\|_{\tilde{\mathcal{A}}}$ can be simplified to
	\begin{align} \label{dual_const}
		\|\mathbf{G}\|_{\tilde{\mathcal{A}}} = & \sup_{\|\mathbf{U}\|_{\mathcal{A}} \leq 1} \langle\mathbf{U}, \mathbf{G}\rangle \notag \\
		=& \sup_{\|\mathbf{U}\|_{\mathcal{A}} \leq 1} \Re\left\{ \operatorname{Tr}(\mathbf{G}^H \mathbf{U})\right\} \notag \\
		=& \sup_{\substack{  c_i >0, \theta_i \in\left(-\frac{\pi}{2}, \frac{\pi}{2}\right],  \\ \varphi_i \in[0,2 \pi),\sum_i c_i \leq 1 }} \Re \left\{ \operatorname{Tr} \sum_i  c_i e^{j \varphi_i} \mathbf{G}^H  \mathbf{b}(\theta_i) \mathbf{a}_r^H(\overline{\theta}_i)  \right\} \notag \\
		=& \sup_{\substack{   \theta_i \in\left(-\frac{\pi}{2}, \frac{\pi}{2}\right],  \\ c_i >0,\sum_i c_i \leq 1 }} \sum_i  c_i \Re \left\{ \operatorname{Tr} \left(\mathbf{G}^H  \mathbf{b}(\theta_i) \mathbf{a}_r^H(\overline{\theta}_i) \right)  \right\} \notag \\
		=& \sup_{\theta_i \in\left(-\frac{\pi}{2}, \frac{\pi}{2}\right]} \left| \operatorname{Tr}(\mathbf{G}^H  \mathbf{b}(\theta) \mathbf{a}_r^H(\overline{\theta})) \right| \notag \\
		=& \sup_{\theta_i \in\left(-\frac{\pi}{2}, \frac{\pi}{2}\right]} \left\|\mathbf{G}^H  \mathbf{b}(\theta) \mathbf{a}_r^H(\overline{\theta}) \right\|_F^2.
	\end{align}
	Next, we construct a Hermitian matrix  $\begin{bmatrix}
		\mathbf{W} & \mathbf{G} \\
		\mathbf{G}^H & \varrho \mathbf{I}_N
	\end{bmatrix}$, where $\varrho \in \mathbb{R}$ is a hyperparameter and $\mathbf{W} \in \mathbb{C}^{M\times M}$ is a  Hermitian matrix. This matrix is   semi-definite positive, i.e., \begin{equation}  \label{const1}
		\begin{bmatrix}
			\mathbf{W} & \mathbf{G} \\
			\mathbf{G}^H & \varrho\mathbf{I}_N
		\end{bmatrix} \succeq 0,
	\end{equation} 
	if and only if
	\begin{align}
		&\mathbf{W} \succeq 0, \label{const2}\\
		&\mathbf{W} - \varrho^{-1}\mathbf{G}\mathbf{G}^H \succeq 0.
	\end{align}
	Therefore, for any given matrix $\boldsymbol{\Psi} \in \mathbb{C}^{M\times N}$, we have 
	\begin{equation}
		\operatorname{Tr} \left(\boldsymbol{\Psi}^H (\mathbf{W} - \varrho^{-1}\mathbf{G}\mathbf{G}^H)\boldsymbol{\Psi}\right) \geq 0,
	\end{equation}
	yielding
	\begin{equation}  
		\|\mathbf{G}^H \boldsymbol{\Psi}\|_F^2 \leq \varrho \operatorname{Tr}(\boldsymbol{\Psi}^H \mathbf{W} \boldsymbol{\Psi}).
	\end{equation}
	Let $\boldsymbol{\Psi}=\mathbf{b}(\theta) \mathbf{a}_r^H(\overline{\theta})$, we have
	\begin{align} \label{psi_exp}
		\|\mathbf{G}^H \mathbf{b}(\theta) \mathbf{a}_r^H(\overline{\theta})\|_F^2 \leq \varrho \operatorname{Tr} \left(\mathbf{a}_r(\overline{\theta}) \mathbf{b}^H(\theta) \mathbf{W} \mathbf{b}(\theta) \mathbf{a}_r^H(\overline{\theta}) \right) \notag \\
		= \varrho N\operatorname{Tr}(\mathbf{W}) + \varrho N \sum_{v\neq 0} e^{j\pi \sin(\theta)} \sum_m W_{m,m+v}.
	\end{align}
	For the matrix $\mathbf{W}$, if we have
	\begin{align}
		&\operatorname{Tr}(\mathbf{W}) = \beta^2/(\varrho N), \label{const3} \\
		&\sum_m W_{m,m+v} = 0, v\neq 0, \label{const4}
	\end{align}
	the expression \eqref{psi_exp} indicates that
	\begin{equation}
		\|\mathbf{G}^H \mathbf{b}(\theta) \mathbf{a}_r^H(\overline{\theta})\|_F^2 \leq \beta^2.
	\end{equation}
	Finally, the original problem \eqref{prob_ori} can be transformed into 
	\begin{subequations} \label{prob_opt}
		\begin{align}
			\min \limits_{ \mathbf{G}, \mathbf{W}} \quad & \operatorname{Tr}\left[(\mathbf{Y}\mathbf{D}^H- \mathbf{G})(\mathbf{D} \mathbf{D}^H)^{-1} \left( \mathbf{Y}\mathbf{D}^H- \mathbf{G}\right)^H \right] \\
			\text { s.t. } \quad & \eqref{const1}, \eqref{const2}, \eqref{const3}, \eqref{const4}.
		\end{align}
	\end{subequations}
	The  problem above is convex and can be efficiently solved by   convex optimization toolbox, such as the CVX toolbox. The computational complexity primarily  depends on the size of the semi-definite positive matrix in \eqref{const1}, resulting in an approximated complexity of $\mathcal{O}\left((N+M)^{3.5}\right)$ \cite{ZHANG201995}.
	Denoting the optimal solution in \eqref{prob_opt} as $\hat{\mathbf{G}}$, the DoAs can be obtained by identifying the $K$ largest peak values of $f(\theta)$ defined below according to the dual constraint in \eqref{dual_const},
	\begin{equation}
		f(\theta) \triangleq |\mathbf{a}_r^H(\overline{\theta}) \hat{\mathbf{G}}^H \mathbf{b}(\theta)|, \forall \theta \in \left(-\frac{\pi}{2}, \frac{\pi}{2}\right].
	\end{equation}
	
	\section{CRB for DoA Estimation}
	The CRB serves as a lower bound on the variance of any unbiased estimator. In this section, we would like to explore the performance limitation of DoA estimation based on the IRS-assisted sensing model in \eqref{Y_re}.  
	
	Let $\boldsymbol{\xi} \triangleq [\theta_1, \ldots, \theta_K, \overline{\alpha}_1, \ldots, \overline{\alpha}_K]^T \in \mathbb{R}^{3K\times 1} $ denote the vector of the unknown targets' parameters, where $\overline{\boldsymbol{\alpha}}_k \triangleq [\Re\{\alpha_k\}, \Im\{\alpha_k\}]$. By vectorizing \eqref{Y_re}, we have
	\begin{equation}
		\operatorname{vec}(\mathbf{Y}) = \operatorname{vec}(\boldsymbol{\Upsilon}) + \operatorname{vec}(\mathbf{N}),
	\end{equation}
	where $\boldsymbol{\Upsilon}=\mathbf{B}(\boldsymbol{\theta}) \boldsymbol{\Lambda} \mathbf{Q}^H (\boldsymbol{\theta}) \mathbf{D}$. Let $\mathbf{F}\in \mathbb{R}^{3K\times 3K}$ denote the Fisher information matrix for estimating $\boldsymbol{\xi}$. Since $\operatorname{vec}(\mathbf{Y}) \sim \mathcal{CN}(\operatorname{vec}(\boldsymbol{\Upsilon}), \sigma^2 \mathbf{I}_{ML})$, each entry of $\mathbf{F}$ is given by \cite{kay1993fundamentals}
	\begin{align}
		\mathbf{F}_{i,\ell} =& \frac{2}{\sigma^2} \Re\left\{\frac{\partial \left(\operatorname{vec}(\boldsymbol{\Upsilon})\right)^H} {\partial \xi_i} \frac{\partial \operatorname{vec}(\boldsymbol{\Upsilon})} {\partial \xi_\ell}\right\} \notag \\
		\overset{(a)}{=}& \frac{2}{\sigma^2} \Re \operatorname{Tr} \left\{\frac{\partial \boldsymbol{\Upsilon}^H} {\partial \xi_i} \frac{\partial \boldsymbol{\Upsilon}} {\partial \xi_\ell} \right\}, \forall i,\ell,
	\end{align}
	where $(a)$ holds due to $\operatorname{Tr} (\boldsymbol{\Upsilon}^H \boldsymbol{\Upsilon}) = \left(\operatorname{vec}(\boldsymbol{\Upsilon})\right)^H \operatorname{vec}(\boldsymbol{\Upsilon})$. Let $\mathbf{e}_i$ denote the $i$-th column of the identity matrix, we have
	\begin{align}
		\frac{\partial \boldsymbol{\Upsilon}} {\partial \theta_i} = \dot{\mathbf{B}}\mathbf{e}_i \mathbf{e}_i^H \boldsymbol{\Lambda} \mathbf{Q}^H \mathbf{D} + \mathbf{B} \boldsymbol{\Lambda} \mathbf{e}_i \mathbf{e}_i^H  \dot{\mathbf{Q}}^H \mathbf{D},
	\end{align}
	where 
	$
	\dot{\mathbf{B}} \triangleq \left[\frac{\partial \mathbf{b}(\theta_1)} {\partial \theta_1}, \ldots, \frac{\partial \mathbf{b}(\theta_K)} {\partial \theta_K} \right] $, $
	\dot{\mathbf{Q}} \triangleq \left[\frac{\partial \mathbf{a}_r(\overline{\theta}_1)} {\partial \theta_1}, \ldots, \frac{\partial \mathbf{a}_r(\overline{\theta}_K)} {\partial \theta_K} \right]$.
	Note that
	\begin{align}
		\operatorname{Tr}\left( \left(\dot{\mathbf{B}}\mathbf{e}_i \mathbf{e}_i^H \boldsymbol{\Lambda} \mathbf{Q}^H \mathbf{D}\right)^H \left(\dot{\mathbf{B}}\mathbf{e}_\ell \mathbf{e}_\ell^H \boldsymbol{\Lambda} \mathbf{Q}^H \mathbf{D}\right) \right) \notag \\
		= \mathbf{e}_i^H \left(\dot{\mathbf{B}}^H \dot{\mathbf{B}}\right) \mathbf{e}_\ell \mathbf{e}_\ell^H  \left(\boldsymbol{\Lambda} \mathbf{Q}^H \mathbf{D} \mathbf{D}^H \mathbf{Q} \boldsymbol{\Lambda}^H \right) \mathbf{e}_i.
	\end{align}
	Consequently, by letting $\mathbf{R}_D \triangleq \mathbf{D} \mathbf{D}^H$ denote the covariance matrix of $\mathbf{D}$, we have
	\begin{align} \label{crb_theta_theta}
		\operatorname{Tr} \left\{\frac{\partial \boldsymbol{\Upsilon}^H} {\partial \boldsymbol{\theta}} \frac{\partial \boldsymbol{\Upsilon}} {\partial \boldsymbol{\theta}}\right\} =& \left(\dot{\mathbf{B}}^H \dot{\mathbf{B}}\right) \odot \left(\boldsymbol{\Lambda} \mathbf{Q}^H \mathbf{R}_D \mathbf{Q} \boldsymbol{\Lambda}^H\right)^T \notag \\
		&+ \left(\dot{\mathbf{B}}^H {\mathbf{B}}\right) \odot \left(\boldsymbol{\Lambda} \dot{\mathbf{Q}}^H \mathbf{R}_D \mathbf{Q} \boldsymbol{\Lambda}^H\right)^T \notag \\
		&+ \left( {\mathbf{B}}^H \dot{\mathbf{B}}\right) \odot \left(\boldsymbol{\Lambda} \mathbf{Q}^H \mathbf{R}_D \dot{\mathbf{Q}} \boldsymbol{\Lambda}^H\right)^T \notag \\
		&+ \left( {\mathbf{B}}^H  {\mathbf{B}}\right) \odot \left(\boldsymbol{\Lambda} \dot{\mathbf{Q}}^H \mathbf{R}_D \dot{\mathbf{Q}} \boldsymbol{\Lambda}^H\right)^T .
	\end{align}
	Similarly, by letting $\boldsymbol{\alpha} \triangleq [\overline{\boldsymbol{\alpha}}_1, \ldots, \overline{\boldsymbol{\alpha}}_K]^T \in \mathbb{R}^{2K\times 1}$, we have
	\begin{align}
		\frac{\partial \boldsymbol{\Upsilon}} {\partial \overline{\boldsymbol{\alpha}}_k} = [1,j] \otimes \left( \sqrt{N_b P_t} \alpha_g \mathbf{B} \mathbf{e}_k \mathbf{e}_k^H \mathbf{Q}^H   \mathbf{D} \right),
	\end{align}
	\begin{small}
		\begin{align} 
			\operatorname{Tr} \left\{\frac{\partial \boldsymbol{\Upsilon}^H} {\partial \boldsymbol{\alpha}} \frac{\partial \boldsymbol{\Upsilon}} {\partial \boldsymbol{\alpha}}\right\} =& N_b P_t |\alpha_g|^2 \mathbf{I}_2 \otimes \left(\left(  {\mathbf{B}}^H  {\mathbf{B}}\right) \odot \left(  \mathbf{Q}^H \mathbf{R}_D \mathbf{Q}  \right)^T\right) , \label{crb_alpha_alpha} \\
			\operatorname{Tr} \left\{\frac{\partial \boldsymbol{\Upsilon}^H} {\partial \boldsymbol{\theta}} \frac{\partial \boldsymbol{\Upsilon}} {\partial \boldsymbol{\alpha}}\right\} = &\sqrt{N_b P_t} \alpha_g [1,j] \otimes \left( \left(\dot{\mathbf{B}}^H {\mathbf{B}}\right) \odot \left(  \mathbf{Q}^H \mathbf{R}_D \mathbf{Q} \boldsymbol{\Lambda}^H \right)^T \right. \notag \\
			&+ \left( {\mathbf{B}}^H {\mathbf{B}}\right) \odot \left(  \mathbf{Q}^H \mathbf{R}_D \dot{\mathbf{Q}} \boldsymbol{\Lambda}^H \right)^T \bigg). \label{crb_theta_alpha}
		\end{align}
	\end{small}
	Thus, the Fisher information matrix can be expressed as 
	\begin{equation}
		\mathbf{F} = \frac{2}{\sigma^2} \begin{bmatrix}
			\Re \operatorname{Tr} \left\{\frac{\partial \boldsymbol{\Upsilon}^H} {\partial \boldsymbol{\theta}} \frac{\partial \boldsymbol{\Upsilon}} {\partial \boldsymbol{\theta}}\right\} & \Re \operatorname{Tr} \left\{\frac{\partial \boldsymbol{\Upsilon}^H} {\partial \boldsymbol{\theta}} \frac{\partial \boldsymbol{\Upsilon}} {\partial \boldsymbol{\alpha}}\right\} \\
			\Re \left\{\operatorname{Tr} \left\{\frac{\partial \boldsymbol{\Upsilon}^H} {\partial \boldsymbol{\theta}} \frac{\partial \boldsymbol{\Upsilon}} {\partial \boldsymbol{\alpha}}\right\}\right\}^T & \Re \operatorname{Tr} \left\{\frac{\partial \boldsymbol{\Upsilon}^H} {\partial \boldsymbol{\alpha}} \frac{\partial \boldsymbol{\Upsilon}} {\partial \boldsymbol{\alpha}}\right\}
		\end{bmatrix}.
	\end{equation}
	Therefore, the CRB of the $k$-th DoA estimation is given by
	\begin{equation}
		\mathrm{CRB}(\theta_k) = \operatorname{var}\{\theta_k\} = \left[\mathbf{F}^{-1}\right]_{k,k},
	\end{equation}
	where $\left[\mathbf{F}^{-1}\right]_{k,k}$ denotes the $k$-th diagonal entry of the matrix $\mathbf{F}^{-1}$. In the case of a single target and $\mathbf{R}_D = L \mathbf{I}_N$, the CRB for DoA estimation can be simplified as \cite{10304548}
	\begin{equation} \label{crb_onetarget}
		\mathrm{CRB}(\theta_0) = \frac{6\sigma^2} {L N_b P_t |\alpha_g|^2 |\alpha_0|^2 \pi^2 \cos^2(\theta_0) MN(M^2 + N^2 -2)}.
	\end{equation}
	It is observed from \eqref{crb_onetarget} that the CRB is inversely proportional to $M N^3+ N M^3$. This relationship can be elucidated as follows.  When $\mathbf{R}_D = L \mathbf{I}_N$, the IRS REs sense the entire space, analogous to the IRS SEs that can receive the signals from the entire space. It is also noteworthy  that $\mathbf{B}$ and $\mathbf{Q}$ in \eqref{crb_theta_theta}, \eqref{crb_alpha_alpha} and \eqref{crb_theta_alpha} are interchangeable with $\mathbf{R}_D = L \mathbf{I}_N$, indicating the equivalent roles of IRS REs and SEs.  Let us take a closer look at $\mathcal{O}(N M^3)$. Note that the IRS SEs inherently provide a spatial direction gain of $\mathcal{O}(M^3)$ \cite{book_loc}. If the IRS REs are consistently  directed towards a specific angle during the entire $L$ time slots, a beamforming gain of $\mathcal{O}(N^2)$ can be achieved \cite{9122596,10284917}. However, in order to sense the targets, the IRS REs should adjust the reflection coefficients in each time slot, thus  contributing only  a gain of $\mathcal{O}(N)$. Therefore, the combined gain totals $\mathcal{O}(N M^3)$, and the CRB exhibits an inverse proportionality to $M N^3+ N M^3$. When there are multiple targets, the interference among them implies that the CRB for one target serves as a lower bound for that of multiple targets, i.e., $\mathrm{CRB}(\theta_0) \leq \mathrm{CRB}(\theta_k)$.
	\begin{figure}[t]
		\vspace{-0.1cm}
		\begin{centering}
			\includegraphics[width=2.45in, height=1.9in]{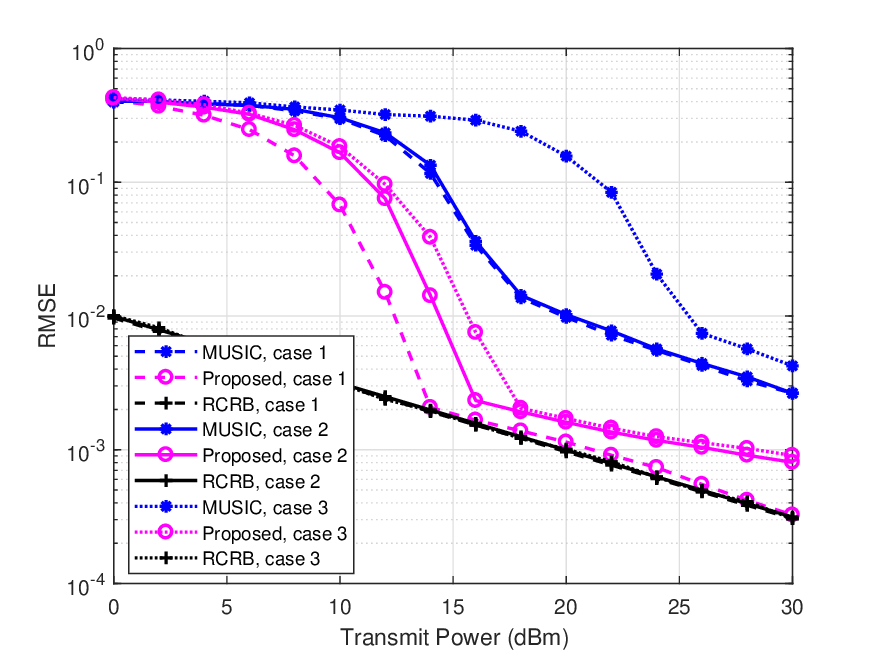}\caption{RMSE versus $P_t$.}\label{fig_pt}
		\end{centering}
		\vspace{-0.5cm}
	\end{figure}
	
	\section{Numerical Results}
	In this section, numerical experiments are conducted to evaluate the performance of the proposed ANM-based DoA estimation method. The carrier frequency is set to $f_c = 28$ GHz, and the transmit signal is $s(t) = 1, \forall t$. Other system parameters are configured as follows unless specified otherwise: $N_b=64$, $N=64$, $M=8$, $L=N$, $\vartheta_{BI}=-60^\circ$, $d_{BI}=30$ m, $K=3$, $\boldsymbol{\theta}= [-10^\circ, 10^\circ, 30^\circ]^T$, $d_k = 5$ m, $\kappa_k=10$ dBsm, $\sigma^2=-120$ dBm, $\varrho = \beta^2/N = 1000$, and $\mathbf{D}$ is set as the DFT matrix. The root-MSE (RMSE) is defined as $\mathrm{RMSE} \triangleq \sqrt{\frac{1}{K N_\text{NS}} \sum_{i=1}^{N_\text{NS}}\|\boldsymbol{\theta} - \hat{\boldsymbol{\theta}}_i\|^2}$, where $N_\text{NS}$ is the number of Monte Carlo trials, and $\hat{\boldsymbol{\theta}}_i$ denotes the estimated DoA in the $i$-th trial. The root-CRB (RCRB) is defined as $\mathrm{RCRB} \triangleq \sqrt{\frac{1}{K}\sum_{k=1}^K \mathrm{CRB}(\theta_k) }$. The benchmark is the MUSIC-based method \cite{10284917} labeled ``MUSIC".
	
	\begin{figure*}[t] 
		\begin{minipage}[t]{0.33\linewidth}
			\centering
			\includegraphics[width=2.45in, height=1.9in]{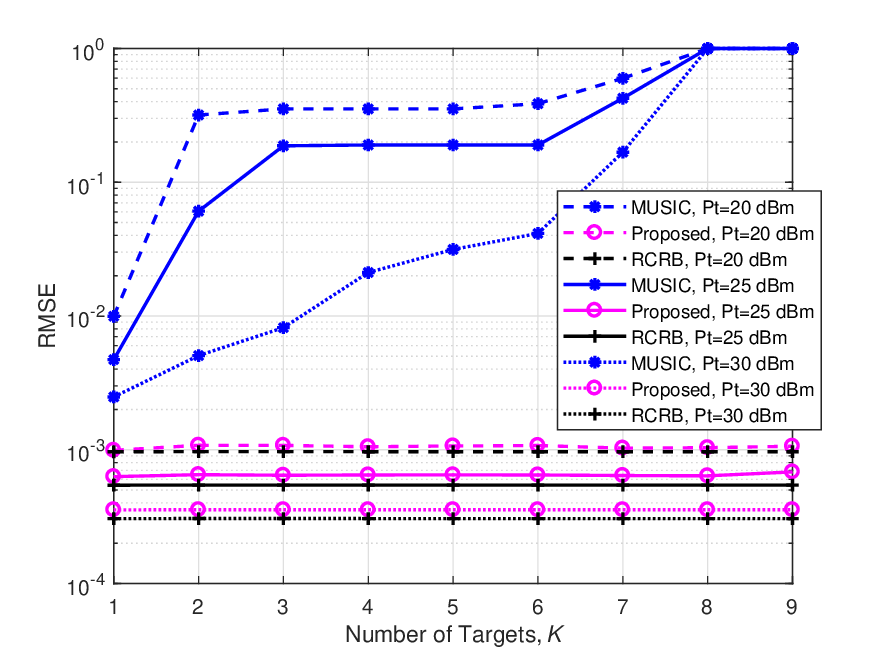}
			\caption{ RMSE versus $K$.} \label{fig_K}
		\end{minipage}%
		\hspace{0 cm}
		\begin{minipage}[t]{0.33\linewidth}
			\centering
			\includegraphics[width=2.45in, height=1.9in]{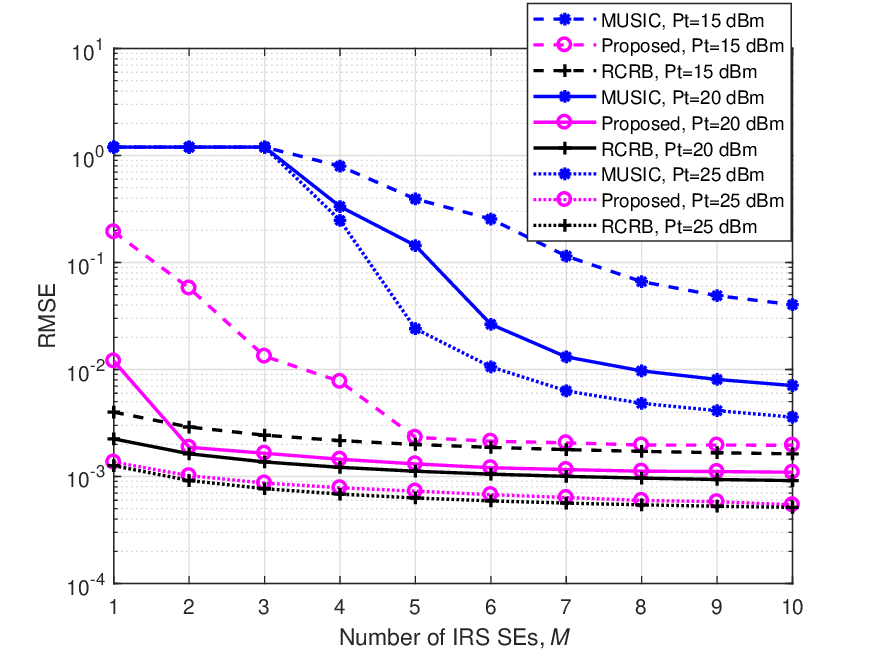}
			\caption{ {RMSE versus $M$ ($N+M=72$).}}\label{fig_M}
		\end{minipage}
		\hspace{0 cm}
		\begin{minipage}[t]{0.33\linewidth}
			\centering
			\includegraphics[width=2.45in, height=1.9in]{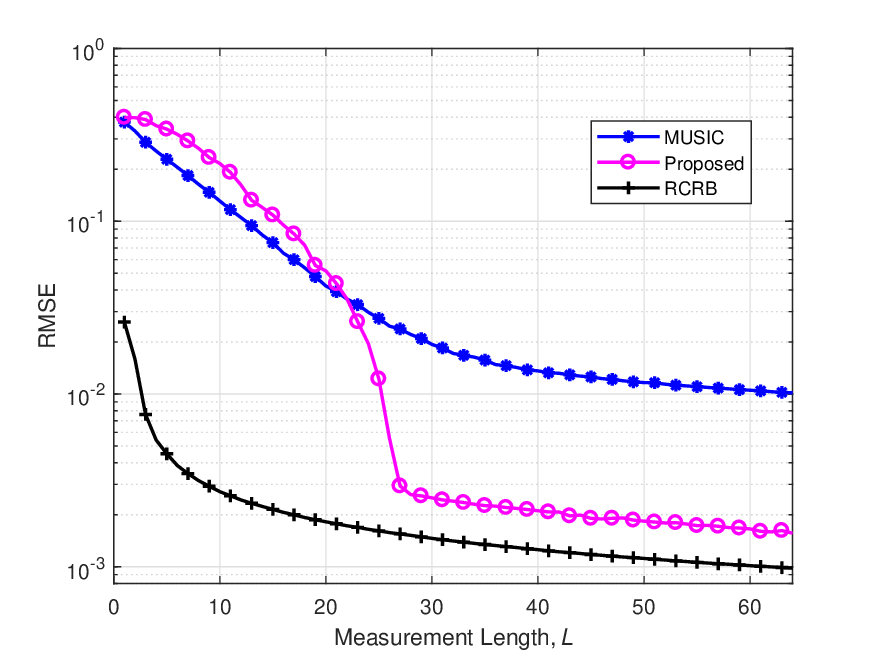}
			\caption{RMSE versus $L$ ($P_t = 20$ dBm).}\label{fig_L}
		\end{minipage}		
		\vspace{-0.7cm}
	\end{figure*}
	
	First, Fig. \ref{fig_pt} illustrates the RMSE versus the transmit power $P_t$ for three cases: 1) Case 1: $\boldsymbol{\theta}= [-10^\circ, 10^\circ, 30^\circ]^T$ and $\mathbf{D} = \mathrm{DFT}$; 2) Case 2: $\boldsymbol{\theta}= [-10^\circ, 10^\circ, 30^\circ]^T$, and the phase shifts of $\mathbf{D}$ follow a uniform distribution over $[0, 2\pi)$; 3) Case 3: $\boldsymbol{\theta}= [-10^\circ, 10^\circ, 20^\circ]^T$, and the phase shifts of $\mathbf{D}$ follow a uniform distribution over $[0, 2\pi)$. Several interesting findings emerge. First, the proposed ANM-based method outperforms the MUSIC-based method in all cases, as it fully utilizes the active and passive element resources (i.e., $\mathbf{B}(\boldsymbol{\theta})$ and $\mathbf{Q}(\boldsymbol{\theta})$), whereas the MUSIC method only utilizes the receive SEs (i.e., $\mathbf{B}(\boldsymbol{\theta})$). {Second, the proposed method nearly reaches the performance limit when employing the DFT measurement matrix. This phenomenon may be attributed to the following reason. The DFT matrix results in an identity  covariance matrix, implying that each target receives equivalent power from the CRB perspective. As a result, the proposed method effectively estimates DoAs. However, with a random measurement matrix, certain targets receive more power while others receive less, intensifying  their interference and diminishing estimation performance. Consequently, the results deviate significantly from the CRB.}  Third, as for case 3 where two targets are close to each other, the performance of MUSIC degrades dramatically, while the performance of the proposed method varies little. This is because the resolution of the proposed method, which  mainly depends on the number of IRS REs $N$, outperforms the MUSIC method, which mainly depends on the number of IRS SEs $M$.
	
	We then investigate the RMSE versus the number of targets $K$ in Fig. \ref{fig_K} considering various transmit power levels. The target angles are taken from the set $\{\pm60^\circ, \pm45^\circ,\pm30^\circ,\pm15^\circ, 0^\circ\}$. The MUSIC method performs well in the case of one target, but experiences a dramatic performance drop as $K$ increases due to its limited resolution.  It fails when $K \geq M=8$. In contrast, the proposed method performs well even when $K=9$, since $N> K$ and it utilizes both $\mathbf{B}(\boldsymbol{\theta})$ and $\mathbf{Q}(\boldsymbol{\theta})$ for DoA estimation.
	
	Next, Fig. \ref{fig_M} displays the RMSE versus the number of IRS SEs $M$ when $K=3$ and $N+M=72$ under different transmit power levels. The DoA estimation accuracy increases as $M$ grows since more echo power can be obtained. The proposed method can even work when $M=1$ since it can also utilize $\mathbf{Q}(\boldsymbol{\theta})$ to sense the targets. However, its performance is limited when the transmit power is low. Therefore, a moderate number of IRS SEs  should be chosen to balance the hardware cost and sensing performance.

	Finally, we explore the RMSE with respect to the measurement length $L$ in Fig. \ref{fig_L}, where $P_t=20$ dBm and the phase shifts of $\mathbf{D}$ follow a uniform distribution over $[0, 2\pi)$. The proposed method performs slightly worse than MUSIC  when $L<23$, likely due to limited measurements. However, when $L>30$, the performance of the proposed method tends to stabilize and significantly outperforms  MUSIC. Therefore, the proposed method is likely to operate with moderate measurements.
	
	\section{Conclusion}
	This work investigates a multi-target DoA estimation problem in a semi-passive IRS-assisted localization system. {The ANM method is adopted to fully extract the DoA information embedded in the IRS REs and SEs matrices}. The CRB for DoA estimation is then derived to evaluate the sensing performance. Simulation results have validated the superior accuracy and resolution performance of the proposed ANM-based DoA estimation method over  MUSIC. Future work may consider  efficient DoA estimation approaches under few IRS measurements.
	
	\bibliographystyle{IEEEtran}
	\bibliography{anm.bib}
	
\end{document}